\documentclass{article}
\usepackage{ijcai22}

\usepackage{subfigure}
\usepackage{float}
\usepackage{makecell}
\usepackage{times}

\usepackage{soul}
\usepackage{url}
\usepackage[hidelinks]{hyperref}
\usepackage[utf8]{inputenc}
\usepackage[small]{caption}
\usepackage{graphicx}
\usepackage{amsmath}
\usepackage{booktabs}
\title{A Decentralized Communication Framework based on Dual-Level Recurrence for Multi-Agent Reinforcement Learning}
\author{
Jingchen Li$^1$\and
Haobin Shi$^1$\footnote{Contact Author}\and
Kao-Shing Hwang$^{2}$\\
\affiliations
$^1$School of Computer Science and Engineering of Northwestern Polytechnical University, Xi'an, China\\
$^2$National Sun Yat-Sen University, Kaohsiung, Taiwan\\
\emails
staubs1212@mail.nwpu.edu.com,
shihaobin@nwpu.edu.com,
hwang@ccu.com
}
\begin{document}

\maketitle

\begin{abstract}
We propose a model enabling decentralized multiple agents to share their perception of environment in a fair and adaptive way. In our model, both the current message and historical observation are taken into account, and they are handled in the same recurrent model but in different forms. We present a dual-level recurrent communication framework for multi-agent systems, in which the first recurrence occurs in the communication sequence and is used to transmit communication data among agents, while the second recurrence is based on the time sequence and combines the historical observations for each agent. The developed communication flow separates communication messages from memories but allows agents to share their historical observations by the dual-level recurrence. This design makes agents adapt to changeable communication objects, while the communication results are fair to these agents. We provide a sufficient discussion about our method in both partially observable and fully observable environments. The results of several experiments suggest our method outperforms the existing decentralized communication frameworks and the corresponding centralized training method.
\end{abstract}

\section{Introduction}
Multi-agent reinforcement learning has aroused intense scholarly interest, especially in the fields of robotics~\cite{5,27} and cyber-physical systems~\cite{4}. In some works, multi-agent reinforcement learning is used to solve large-scale decision-making tasks. These works broke a complicated decision-making model down into several sub-decision-making processes, using multi-agent reinforcement learning algorithms to optimize the policies for every process~\cite{6}. With more advanced optimization and deep learning techniques~\cite{25}, multi-agent reinforcement learning enables agents to collaborate with each other and respond to the environment timely, which the traditional machine learning techniques and heuristic methods lack.
\par
The largest challenge in multi-agent reinforcement learning is that the environment becomes unstable for every agent due to the existence of its peers~\cite{7,32}. The main direction for multi-agent reinforcement learning is centralized learning because decentralized multi-agent reinforcement learning lacks the couplings among the behaviors of different agents. Aiming to design more lightweight centralized learning models, some researchers proposed a centralized training and decentralized execution (CTDE) mechanism to train multi-agent system~\cite{12}. This mechanism allows an agent to make decisions according to its own observation or perception, while its peers' behaviors should be considered when evaluating its policy. However, centralized learning is not suitable for distributed control~\cite{28}, especially in large-scale or partially observable multi-agent environments. Communication is mainstream in the case of decentralized training: Sukhbaatar et al.~\cite{9} utilized an additional neural network module to realize continuous communication for multiple agents, aiming at fully cooperative scenarios; Peng et al.~\cite{10} leveraged bidirectional-recurrent network to share latent states for agents, and validated their model on StarCraft testbed; Jiang et al.~\cite{11} introduced attention mechanism into multi-agent communication, in which the attention module is used to judge whether an agent should communicate with others. As a feasible way to share information among agents, communication can enhance policy coordination and make agents gain more decision bases in partially observable environments.
\par
The communication modules for multi-agent can be divided into two types: integration module and recurrent model-based module. The former involves centralized communication, using a centralized network to combine the states for all agents~\cite{13}. For example, Targeted Multi-Agent Communication architecture (TarMAC)~\cite{14} leveraged a soft attention module to conduct a targeted communication behavior, by which every two agents are assigned a communication channel. The latter regards the agent group as a sequence, using recurrent models to process the latent state for the agent sequence. Deep distributed recurrent Q-networks (DDRQN)~\cite{15} is an earlier recurrent model-based method aiming to solve communication-based coordination tasks without any pre-designed communication protocol. Then several improved models are developed, such as CommNet (Communication Net)~\cite{9} and BicNet (Bidirectionally-Coordinated Net)~\cite{10}. Some works also utilized attention mechanisms to enhance recurrent model-based methods. For example, Attentional and Recurrent Message Integration (ARMI) calculated the correlation between the message and the observation by attention mechanism. At the same time, Liu et al.~\cite{17} leveraged a self-attention mechanism to build a communication framework, which can learn both to construct communication groups and decide when to communicate for agents.
\par
The existing communication mechanisms still have limitations. Integration modules result in a massive communication network for a too-large multi-agent system, and it cannot be used in the case of limited communication due to the centralized communication module. As for recurrent model-based communications, the order of agents in the communication sequence may be a hidden danger. Although some works have introduced bidirectional recurrent models into the communication modules, the agents at the two ends of the communication sequence are still hard to send messages. Furthermore, when the communication objects or agent sequence changes, the existing methods have to retrain their communication modules due to the lack of adaptability.
\par
We consider fully cooperative multi-agent systems, and agents can be both homogeneous and heterogeneous. In this work, we propose a dual-recurrent communication model (2ReCom) for multi-agent reinforcement learning. The proposal is a circular recurrent communication module. In our method, agents share messages by the first-level recurrence, while the historical observations are taken into account by the second-level recurrence. Compared with bidirectional recurrent models, our model is fair to all agents in the communication sequence. In 2ReCom, we separate communication messages from memories for every agent so that agents can adapt changeable communication objects in the case of limited communication. With the developed 2ReCom, agents can combine both the current and historical messages, by which the entire environment can be precepted more efficiently.
\par
The main contributions of this work are summarized as follows:
\begin{itemize}
    \item We develop a dual-recurrent communication model for multi-agent reinforcement learning, by which the agents can adapt changeable communication objects while the communication results are fair to all agents.
    \item We propose a gated recurrent network to conduct communication in our model. With that recurrent network, the proposed 2ReCom can guarantee the communication qualities and stabilized communication flow.
    \item We conduct experiments in both partially observable and fully observable environments, and provide sufficient discussions to analyze the experimental results. 
\end{itemize}

\section{The Proposed Method}
In this section, we present a dual-recurrent communication framework (2ReCom). First, we describe the gated recurrent network used in 2ReCom. Then the entire communication framework is given, and we discuss the application of 2ReCom in the case of limited communication.

\subsection{Gated Recurrent Model for Communication}
\begin{figure}
    \centering
    \includegraphics[scale=0.25]{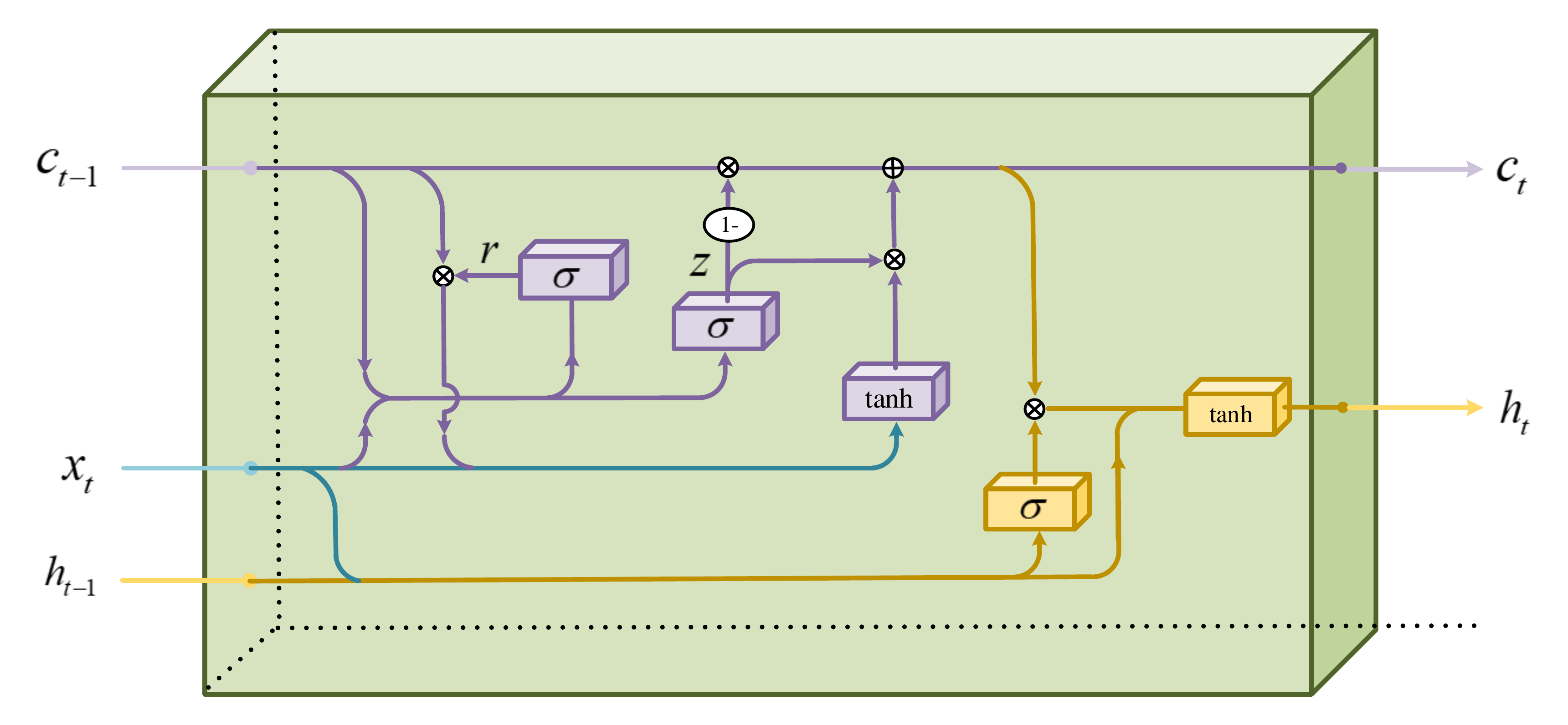}
    \caption{The structure of the gated recurrent network.}
    \label{fig1}
\end{figure}

The entire structure of the proposed gated recurrent network is shown in Fig.~\ref{fig1}. In this section, we just describe the design for the recurrent network. The reason for this design will be presented later after the communication flow is given.
\par
As shown in Fig.~\ref{fig1}, there are three inputs for our gated based recurrent network: the cell state $c_{t-1}$, the current input $x_{t}$, and the communication message $h_{t-1}$. At time $t$, the previous cell state and communication message are fed together with the current input. Because we hope the cell state can remember both the long-term memory and the short-term dependence, we use GRU as the prototype, while the output is separated from the cell state by an additional gated module. Similar to GRU, the cell state is updated by the current input. First, a reset gate processes $c_{t-1}$ and $x_{t}$, calculating the reset information $r$:
\begin{equation}
    r = \text{Sigmoid}(W_{r}\cdot [c_{t-1},x_{t}]),
\end{equation}
where $W_{r}$ denotes the transformation matrix in the reset gate. Then the update gate is used to change the cell state. Let
\begin{equation}
    z = \text{Sigmoid}(W_{z}\cdot [c_{t-1},x_{t}]),
\end{equation}
where $W_{z}$ is the transformation matrix in the update gate. $z$ determines which data should be forgotten in the cell state and which information in the input should be remembered. The new data that should be stored by the cell state is calculated as
\begin{equation}
    \widetilde{c} = \text{tanh}(W_{u}\cdot [r\cdot c_{t-1},x_{t}]),
\end{equation}
where $W_{u}$ is the transformation matrix for the remember. After that, a new cell state is generated:
\begin{equation}
    c_{t} = (1-z)\cdot c_{t-1}+\widetilde{c}.
\end{equation}
\par
It should be noticed that we do not regard the new cell state as the final output of the recurrent model. The additional attention gate, which is the brown part in Fig.~\ref{fig1}, separates the output from the cell state. The attention value is generated through a transformation matrix $W_{a}$ fed by $x_{t}$ and $h_{t-1}$:
\begin{equation}
    a = \text{Softmax}(W_{a}\cdot [x_{t},h_{t-1}]).
\end{equation}
The generated attention value acts on the new cell state. That is, the attention gate determines which data is important in the cell state. We assume the communication message $h_{t-1}$ and cell state $x_{t}$ are in the same feature space, and the attention gate reflects the position for the important data. We will explain why not feed the new cell state with $W_{a}$ later.
\par
The final output is calculated by output gate $W_{o}$:
\begin{equation}
    h_{t} = \text{tanh}(W_{o}\cdot [a\cdot c_{t},h_{t-1}]).
\end{equation}
Because our recurrent network is used to share messages among agents, the output should combine both cell state and communication message. Different from general recurrent models, $h_{t}$ has independence from $c_{t}$, but makes connections with $c_{t}$ through $x_{t}$.
\begin{figure}
    \centering
    \subfigure[]{\includegraphics[scale=0.10]{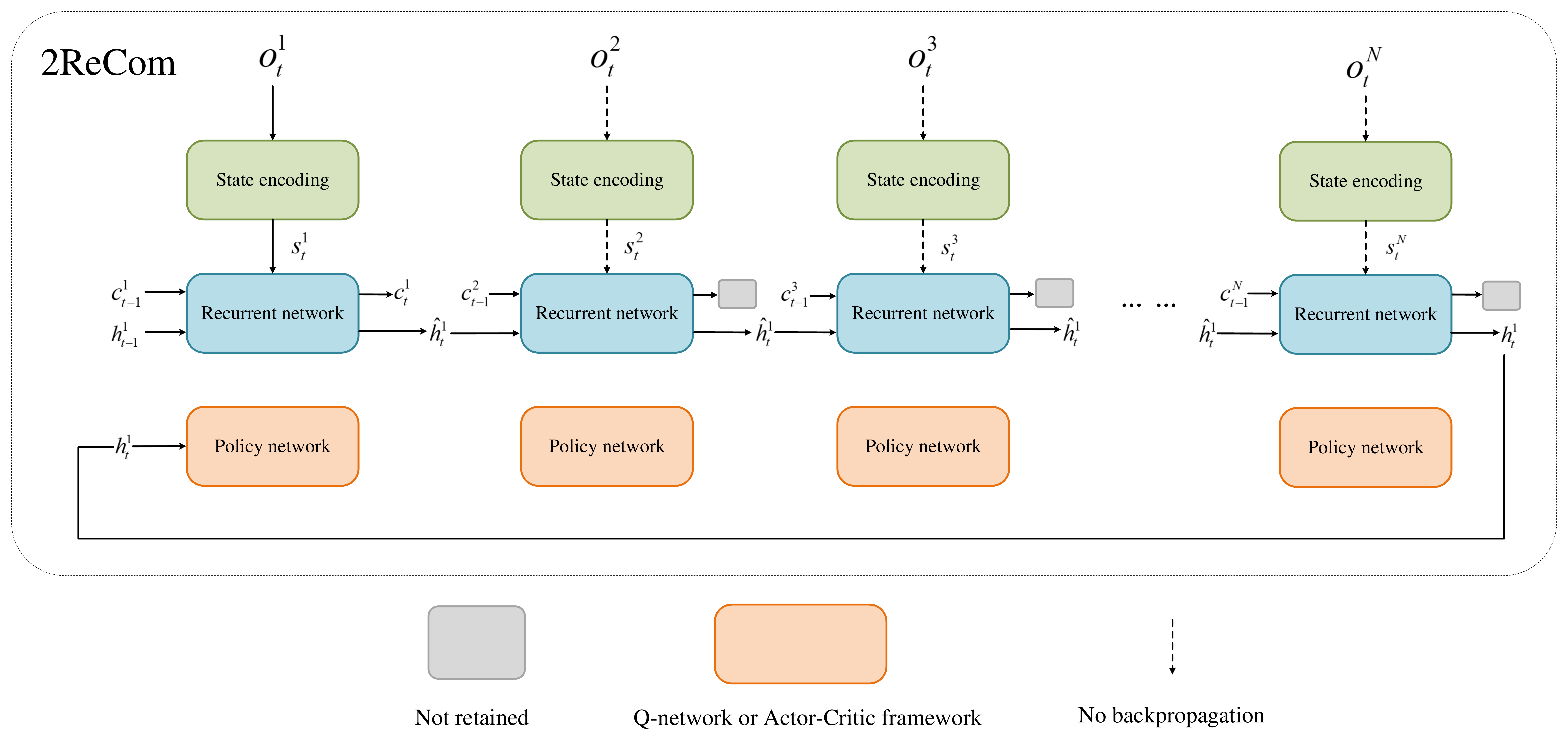}}
    \subfigure[]{\includegraphics[scale=0.08]{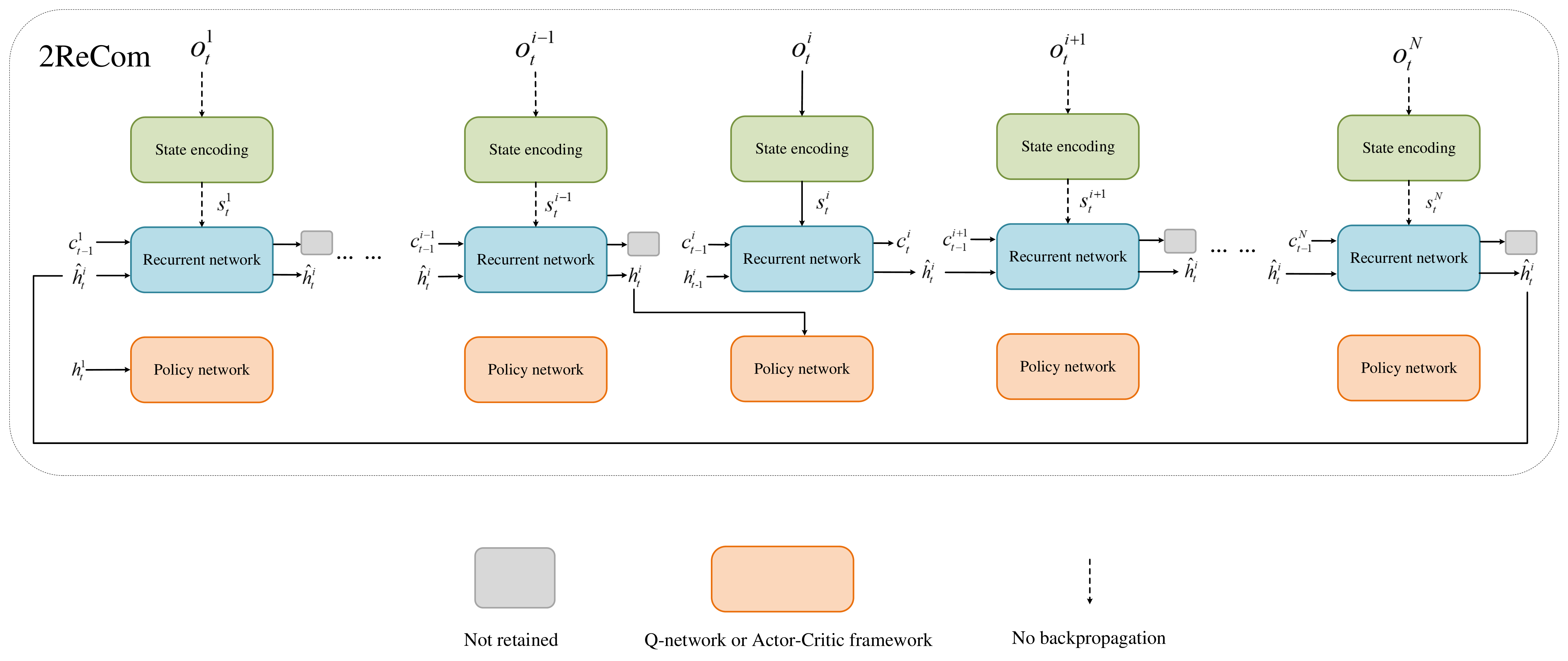}}
    \caption{The 2ReCom framework for $N$ agents. (a) is the communication flow for the first agent, while (b) is the communication flow for the $i$-th agent.}
    \label{fig2}
\end{figure}
\subsection{Dual-Recurrent Communication}
In this section, we present our 2ReCom framework detailedly. In a multi-agent environment with $N$ agents, we use $o^{i}_{t}$ to denote the observation of the $i$-th agent at time $t$. As shown in Fig.~\ref{fig2}, we assign every agent with a state encoding module, a gated recurrent network mentioned before, and a policy network. In our 2ReCom framework, agents share messages through a dual-recurrent model. The state encoding modules transform observations and output hidden states, by which the observations can be mapped into the same feature space. $f(\cdot;\theta ^{r}_{i})$ denotes the recurrent model for the $i$-th agent.
\par
Fig.~\ref{fig2} (a) shows the communication flow for the first agent. After hidden state $s^{1}_{t}$ is calculated, the recurrent model for the first agent updates its cell state and generate a temporary communication vector:
\begin{equation}
    c^{1}_{t},\hat{h}^{1}_{t} = f(c^{1}_{t-1},s^{1}_{t},h^{1}_{t-1};\theta ^{r}_{1}),    
\end{equation}
where $c^{i}_{t}$ is the cell state for the $i$-th agent at time $t$, and $h^{i}_{t}$ denotes the final communication vector for the $i$-th agent. In this process, the new cell state for the first agent $c^{1}_{t}$ is retained and used at the next time step $t+1$. Then the temporary communication vector $\hat{h}^{1}_{t}$ is transmitted to $f(\cdot;\theta ^{r}_{2})$, updating itself by $c^{2}_{t-1}$ and $s^{2}_{t}$:
\begin{equation}
     c^{2}_{t},\hat{h}^{1}_{t}=f(c^{2}_{t-1},s^{2}_{t},\hat{h}^{1}_{t};\theta ^{r}_{2}).
\end{equation}
It should be noticed that in this process, the generated $c^{2}_{t}$ is not retained. The cell state for the $i$-th agent is just updated in the $i$-th agent's communication flow. $\hat{h}^{1}_{t}$ is updated continually through all other agents' recurrent models in this way until the last agent's recurrent model outputs the final communication feature vector for the first agent:
\begin{equation}
    h^{1}_{t} = g^{N}_{t}(g^{N-1}_{t}(\cdots(g^{1}_{t}(h^{1}_{t-1}))\cdots)),
\end{equation}
where $g^{i}_{t}(x) = f(c^{i}_{t-1},s^{i}_{t},x;\theta^{r}_{i})\setminus c^{i}_{t}$.
\par
For the $i$-th agent, its communication flow is shown in Fig.~\ref{fig2} (b). First, a temporary communication vector $\hat{h}^{i}_{t}$ is output by $f(\cdot;\theta^{r}_{i})$, while the new cell state $c^{i}_{t}$ is retained for the communication at the next time step. Then $\hat{h}^{i}_{t}$ is updated continually through the recurrent models of latter agents. After $f(\cdot;\theta^{r}_{N})$ updates $\hat{h}^{i}_{t}$, the temporary communication vector is transmitted to $f(\cdot;\theta^{r}_{1})$. Till $\hat{h}^{i}_{t}$ is transmitted to $f(\cdot;\theta^{r}_{i-1})$, the final communication vector is generated:
\begin{equation}
    h^{i}_{t}=g^{i-1}_{t}(g^{i-2}_{t}(\cdots g^{1}_{t}(g^{N}_{t}(g^{N-1}_{t}(\cdots g^{i}_{t}(h^{i}_{t-1}) \cdots))) \cdots)).
\end{equation}
\par
In a partially observable environment or large-scale multi-agent system~\cite{31}, the limited communication ability should be considered. An agent may have no communication channel with the peers that cannot be observed by it in a partially observable environment. In a large-scale multi-agent system, communicating with all peers is needless for an agent is impossible. So we discuss the application of the developed 2ReCom in the case of limited communication. Because the communication vector is separated from the cell states in the communication flow, 2ReCom can adapt to changeable communication objects. In a partially observable environment, an agent can just communicate with the peers that can be observed by it. Let $N_{i}$ denote the number of the peers observed by the $i$-th agent, the communication flow for the $i$-th agent is:
\begin{equation}
    h^{i}_{t}=g^{j_{N_{i}}}_{t}(g^{j_{N_{i-1}}}_{t}(\cdots g^{i}_{t}(h^{i}_{t-1}) \cdots)), (j_{1},\cdots,j_{N_{i}})\in \widetilde{i}.
\end{equation}
where $\widetilde{i}$ is the set of the observed agents. In a large-scale multi-agent environment, researchers always leveraged approximation to simplify the interaction among agents~\cite{23,24}. When using 2ReCom to train a large-scale multi-agent system, $\widetilde{i}$ can be regarded as the neighbor agents for the $i$-th agent.
\par
Due to the separation of the communication results and cell states, the change of communication objects has no negative effect on the communication results. Even if a new agent becomes a neighbor or observable peer for the $i$-th agent, its cell state contains just the memory of its own historical hidden states, which have no relevance to its historical communications.
\subsection{Interpretation for 2ReCom}
There are two recurrent processes in the developed 2ReCom framework. In the first level recurrence, the hidden states of all agents are regarded as a sequence. For each agent, its own recurrent model first generates a temporary communication vector, and then all other recurrent models update this vector in turn. In this process, the communication vector gains messages from every agents' memory (cell state) by the brown part in Fig.~\ref{fig1}. In the second-level recurrence, the hidden state sequence for each agent is integrated into the cell state. In our gated recurrent model, the communication vector is separated from the cell state, so that the cell states for agents are independent of each other. The cell state contains both long-term memory and short-term dependency instead of dividing them. That is why GRU rather than LSTM is used as the prototype. Moreover, the historical communication vectors of an agent also form a sequence that remains the communication results, which is necessary for the agent to perceive the entire environment.
\par
The gated recurrent model is specially developed for the 2ReCom framework. In our recurrent model, the new cell state is calculated by just the previous cell state and hidden state. In other words, the communication vector does not participate in this process. In the communication flow, just (temporary) communication vector and hidden state co-determine which information in the cell state should be integrated into the communication vector. Due to the new cell state has combined the hidden state and the previous cell state, the communication result is calculated just by itself and the attentive new cell state.   
\par
At time $t$, after all agents execute their actions and get rewards, an experience tuple $<O_{t},O_{t+1},C_{t-1},C_{t},H_{t-1},H_{t},A_{t},R_{t}>$ is stored in replay buffer, where $O_{t}=(o^{1}_{t},o^{2}_{t},\dots,o^{N}_{t})$ is the observations for all agents, $C_{t}=(c^{1}_{t},c^{2}_{t},\dots,c^{N}_{t})$ denotes the cell states, $H_{t}=(h^{1}_{t},h^{2}_{t},\dots,h^{N}_{t})$ is the final communication vectors at time $t$, $A_{t}=(a^{1}_{t},a^{2}_{t},\dots,a^{N}_{t})$ is the joint actions at time $t$, and $R_{t}=(r^{1}_{t},r^{2}_{t},\dots,r^{N}_{t})$ is the rewards vector given by the environment. It should be noticed that in the update process for an agent, all other agents' hidden states do not back-propagate gradients, as shown in Fig.~\ref{fig2}. In a communication flow, the communication vector needs to gain messages through hidden states for all agents. However, there is no coupling between the state encoding modules for other agents and the policy for the current agent, so that the state encoding module for an agent is updated together with just the policy network for it.
\section{Experiment}
In this section, we first conduct experiments in partially observable environments to compare our 2ReCom with several baseline methods. Then we investigate the performances of 2ReCom in a fully observable environment. The Multi-Agent Particle Environment is used as the experimental platform. In these experiments, the policy network in 2ReCom and the baseline methods take DDPG as the prototype. 
\subsection{Baseline Methods}
Five algorithms are used as the baseline methods: ATOC, BicNet, CommNet, MADDPG, DDPG. The first three algorithms are communication-based methods, and MADDPG is the corresponding centralized learning model, while the last one is the independent learning method.
\par
\textbf{ATOC} (Attentional communication model)~\cite{11} designs an attention unit to receive hidden states and action intention for each agent. An agent determines whether to communicate with other agents according to the attention unit. ATOC leverages a bidirectional LSTM unit as the communication channel.
\par
\textbf{BicNet} (Bidirectionally-coordinated net)~\cite{10} uses a bidirectional RNN as the communication channel, allowing agents to share latent states. BicNet provides a vectorized extension for the actor-critic formulation, and it also introduces module sharing to solve the scalability issue.
\par
\textbf{CommNet} (Communication Neural Net)~\cite{9} uses continuous communication to coordinate multi-agent system. CommNet is the typical work to replace manually specified communication protocol with a deep feed-forward network. 
\par
\textbf{MADDPG} (Multi-agent deep deterministic policy gradient)~\cite{12} proposed CTDE mechanism to train multi-agent system, in which the actor-network is decentralized, and the critic network is centralized. Because this work uses DDPG as the reinforcement learning model for 2ReCom, MADDPG can be regarded as the corresponding CTDE method. 
\par
\textbf{DDPG} (Deep deterministic policy gradient) is a decentralized learning method, that is, each agent in DDPG updates its policy independently. 
\subsection{Experiments Settings}
In these experiments, the learning rate is 0.001, the discounted factor is set to 0.99, and the batch size is 1024. Observation for an agent includes its position, its velocity, and the relative locations of other agents or landmarks.
\par
The state encoding module is a linear layer with 64 nodes, and a leaky ReLU is followed for non-linear activation. The output of the gated recurrent model is also 64-dimensional. As for the policy network, the actor-network is a linear layer. The critic-network first encodes the action and messages to two 64-dimensional tensors with leaky ReLU functions, and then the two tensors are concatenated and fed to a linear layer. 
\subsection{Partially Observable Environment}
\subsubsection{Scenarios}
There are two scenarios in the test on partially observable environments: Cooperative Navigation and Predator Prey. Both of them are in a two-dimensional world with continuous space and discrete, and we modify them to partially observable environments. These experiments are used to investigate the performance of 2ReCom in the case of limited communication.
\par
\textbf{Cooperative Navigation} scenario has 20 agents and 20 landmarks. Every agent needs to reach a landmark as soon as possible while colliding is not allowed. In this scenario, each agent has its independent reward. The reward is negatively correlated with the relative distance between the agent and the nearest landmark, and an agent will deserve punishment if it collides with another. At each time step, an agent can observe just 3 nearest agents and 3 nearest landmarks, so that the agents need to communicate with each other to precept the entire environment. 
\par
\textbf{Predator Prey} scenario has 10 predators and 5 targets. The predators need to catch the targets while colliding is not allowed. The speed of targets is twice that of predators, so that predators need to cooperate with each other. In this scenario, we pre-trained the policies for the targets, while our 2ReCom and the baseline methods were used to train the predators. The policies for targets are not updated in the learning processes. At each time step, a predator can observe just 3 nearest peers and 3 nearest targets, so that predators need to communicate with each other for more effective hunting. 
\subsubsection{Results and Analyses}
\begin{figure}
    \centering
    \includegraphics[scale=0.4]{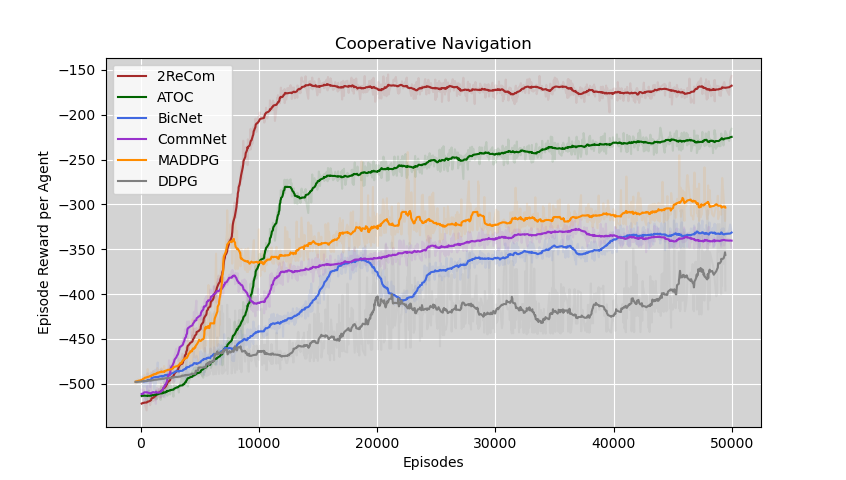}
    \caption{The results on Cooperative Navigation.}
    \label{fig3}
\end{figure}
As shown in Fig.~\ref{fig3}, our 2ReCom achieves the best result on Cooperative Navigation. In a partially observable environment, CTDE-based methods use joint observation and action to evaluate policies, so that the coupling among agents can be captured easily. However, the too-large joint state space and action space result in ineffective learning. The episode reward on MADDPG is just -300, which is far smaller than ATOC and our 2ReCom. ATOC leverages an attention mechanism to conduct communication, by which agents can get suitable communication objects in large-scale environments. The developed 2ReCom adapts changeable communication objects by separating communication messages from cell states, so that 2ReCom can also achieve efficient communication. Moreover, our 2ReCom can combine the communication objects' historical observations by the two-level recurrent model, which is important to build an entire perception in partially observable environments, so that our 2ReCom outperforms ATOC. The other two communication frameworks, BicNet and CommNet, are far behind our 2ReCom because the communication channels in them lack the control of communication objects. DDPG gets the worst performance. It is no doubt that independent learning cannot handle large-scale multi-agent reinforcement learning.
\par
\begin{table}[]
    \centering
    \begin{tabular}{|c|c|c|}
    \hline
         &Episode Reward per Agent& Standard Deviation \\
         \hline
         2ReCom& -170.5 & $\pm 0.04$ \\
         \hline
         ATOC &-228.7 & $\pm 0.07$ \\
         \hline
         BicNet &-334.2 & $\pm 1.29$\\
         \hline
         CommNet & -339.0 & $\pm 0.97$\\
         \hline
         MADDPG & -303.1 &$\pm 0.04$ \\
         \hline
         DDPG &-378.8 &$\pm 8.44$ \\
         \hline
         \end{tabular}
    \caption{The standard deviations on Cooperative Navigation}
    \label{tab2}
\end{table}
Compared with other recurrent model-based communication frameworks, our 2ReCom is fair to all agents due to the novelty communication flow. To validate this, we calculate the standard deviations of the 20 agents' rewards. As shown in Table~\ref{tab2}, 2ReCom gets the best result while achieving a smaller standard deviation. In addition to DDPG, BicNet gets the largest standard deviation. Although the bidirectional recurrent model proves that every agent can receive messages from each other, the communication sequence still results in different communication conditions. In 2ReCom, the communication flows for all agents are similar, and the changeable communication flows guarantee that the uncertain communication objects cannot impact the communication qualities. That is why our 2ReCom is fair to all agents.
\par
\begin{figure}
    \centering
    \includegraphics[scale=0.4]{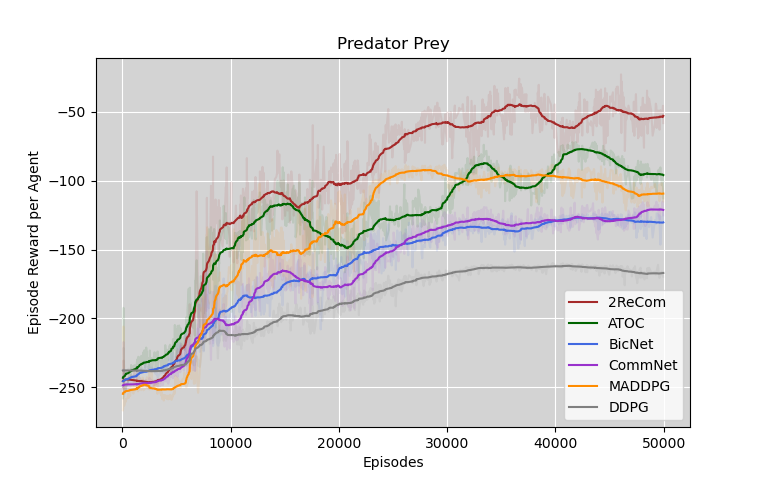}
    \caption{The results on Predator Prey}
    \label{fig5}
\end{figure}
Fig.~\ref{fig5} shows the results on the Predator Prey scenario. The results are consistent with those on Cooperative Navigation: Our 2ReCom outperforms all baseline methods, while other communication frameworks (expect ATOC) fall behind the corresponding CTDE-based method (MADDPG). Because this task has a shared reward for all agents, agents have to learn more advanced cooperation through communication. In our 2ReCom, agents can get messages from the historical states of others, so that the cooperation can be learned more quickly. With the two experiments, the superiority of 2ReCom in the case of limited communication is validated convincingly. 
\par
However, a latent risk is that agents may form several groups spontaneously to avoid communicating with all others. To investigate this phenomenon, we count the communication objects for every agent, judging whether agents can communicate with all peers by changeable communication flows. The result is given in Table~\ref{tab1}, where the line $i$ column $j$ is the number of occurrences of the $j$-th agent within the communication flow for the $i$-th agent. We can know each agent can communicate with most peers instead of fixed communication objects. The communication flows for agents are not symmetrical. For example, the second agent appeared in the communication flow for the first agent three times, but the first agent just appeared four times in the communication flow for the second agent. This is because an agent just communicates with the three nearest peers. If an agent is surrounded by many peers, just three of the peers will appear in its communication flow while it may appear in several peers' communication flows. Our 2ReCom allows agents to adapt changeable communication objects, and this result suggests that agents do not form fixed groups to avoid the change of communication objects. 
\begin{table}[]
    \centering
    \setlength{\tabcolsep}{2mm}{
    \begin{tabular}{|c|cccccccccc|}
        \hline
        Agent & 1 &2 &3 &4&5&6&7&8&9&10 \\
        \hline
         1 & - & 3 & 0 & 17& 12 & 9 & 0& 11 & 19 & 4 \\
         2 & 4 & - & 7 & 1 & 9 & 6 & 21 & 8 & 0 & 19 \\
         3 & 1 & 7 & - & 15 & 11 & 2 & 9 & 14 & 2 & 14 \\
         4 & 15 & 0 & 14 & 8 & 7 & 0 & 2 & 13 & 5 & 11 \\
         5 & 16 & 6 & 9 & 13 & - & 1 & 4 & 16 & 0 & 10 \\
         6 & 11 & 4 & 0 & 2 & 0 & - & 19 & 13 & 17 & 9 \\
         7 & 0 & 22 & 11 & 1 & 5 & 18 & - & 6 & 6 & 2 \\
         8 & 12 & 5 & 14 & 9 & 12 & 17 & 3 & - & 3 & 0 \\
         9 & 17 & 0 & 2 & 4 & 1 & 19 & 8 & 5 & - & 19 \\
         10 & 5 & 16 & 11 & 7 & 8 & 12 & 6 & 0 & 10 & - \\
         \hline
    \end{tabular}}
    \caption{The communication count on Predator Prey.}
    \label{tab1}
\end{table}
\subsection{Fully Observable Environment}
\subsubsection{Scenario}
In this experiment, we compare our 2ReCom with baseline methods on the Cooperative Treasure Collection scenario~\cite{18}. This scenario has two types of agents: hunters and banks. The 6 hunters need to collect treasures and deposit the treasures with the corresponding bank agents. In this scenario, colliding among hunters is not allowable, and all agents can observe the position of each other. We set an individual reward for every agent instead of a shared reward. Compared with the former two scenarios, this scenario requires lower-level cooperation. Because there are just 8 agents in Cooperative Treasure Collection, all agents communicate with each other when trained by 2ReCom. That is, the communication flow for every agent is fixed, and all other agents appear in it.
\subsubsection{Results and Analyses}
\begin{figure}
    \centering
    \includegraphics[scale=0.4]{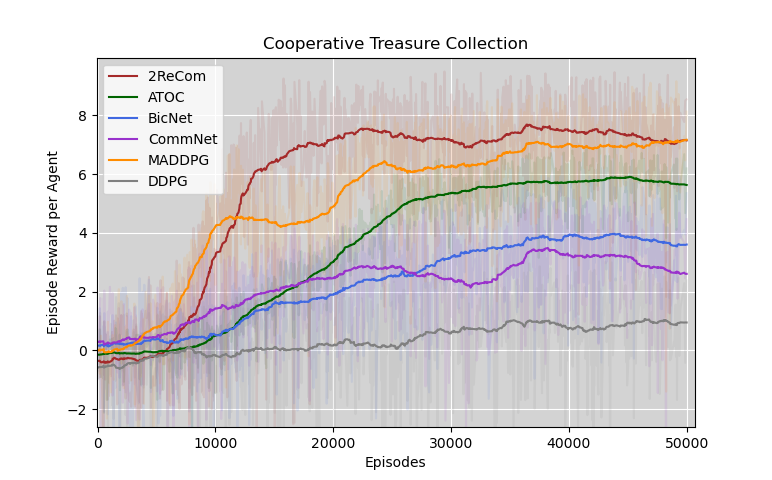}
    \caption{The results on Cooperative Treasure Collection.}
    \label{fig7}
\end{figure}
The results on Cooperative Treasure Collection are given in Fig.~\ref{fig7}. Our 2ReCom is the only decentralized communication method that outperforms MADDPG. The mean episode reward for 2ReCom reaches 7.3, while those for other communication frameworks are less than 6.0. Although the state-of-the-art centralized method~\cite{18} got a larger score in this scenario, this experiment still suggests the developed 2ReCom is better than other decentralized communication methods in fully observable environments, and it even rivals the corresponding centralized learning model (MADDPG). In this experiment, an agent communicates with all others at every time step. The fixed communication flows make us investigate whether the sequence of agents impacts the communication results in 2ReCom. In a heterogeneous multi-agent system, the perceptions of agents are different, so that the risk of totally different communication results may exist when the sequence of agents is changed. To validate that, we conducted this experiment several times with different agent sequences to investigate whether 2ReCom gets different results. The results are shown in Table~\ref{tab3}. On all the four control groups, the results for our 2ReCom are about 7.3, which suggests that the efficiency of our 2ReCom is not impacted by the agent sequence in the case of heterogeneous multi-agent systems. Conversely, when the agent sequence changes, BicNet gets a very different result. In our opinion, the communication messages extracted by gated recurrent models are more effective than those extracted by RNN. The gated recurrent model specially designed by us strengthens this superiority.  
\begin{table}[]
    \centering
    \setlength{\tabcolsep}{1.3mm}{
    \begin{tabular}{|c|c|c|c|c|}
    \hline
         sequence & sequence 1&sequence 2&sequence 3&sequence 4  \\
         \hline
         2ReCom& 7.319 & 7.326&7.292 &7.391 \\
         \hline
         BicNet& 3.977 & 3.016 &3.928 &3.444\\
         \hline
    \end{tabular}}
    \caption{The results on different agent sequences}
    \label{tab3}
\end{table}

\section{Conclusion}
In this work, we develop a new communication framework for decentralized multi-agent reinforcement learning. Our 2ReCom has two main superiorities: We regard historical states of agents as a part of communication information, proposing a dual-recurrence for decentralized multi-agent systems; the developed 2ReCom separates communications from memories, making agents adapt to changeable communication objects. We analyze applications of the proposed 2ReCom in the case of different multi-agent systems, and a sufficient discussion about module sharing is provided. Compared with other communication frameworks, our 2ReCom is fair to all agents, and the agent sequence makes no impact on the communication results. The experiments on both partially and fully observable environments proved that our 2ReCom is better than the existing communication frameworks and the corresponding centralized learning method.

\bibliographystyle{named}
\bibliography{ijcai22}

\end{document}